\documentclass{article}

\usepackage{epsfig}
\usepackage{rotating}

\newcommand{\ba}{\begin{eqnarray}}
\newcommand{\ea}{\end{eqnarray}}
\begin{document}

\title{Symmetries in physics 
\footnote{Lecture notes: XIII Escuela de Verano en F{\'{\i}}sica, 
M\'exico, D.F., M\'exico, August 9 - 19, 2005}}
\author{Roelof Bijker\\
ICN-UNAM, AP 70-543, 04510 M\'exico, DF, M\'exico \\
E-mail: bijker@nucleares.unam.mx}
\date{August 24, 2005}
\maketitle

\begin{abstract}
The concept of symmetries in physics is briefly reviewed. 
In the first part of these lecture notes, some of the basic 
mathematical tools needed for the understanding of symmetries 
in nature are presented, namely group theory, Lie groups and 
Lie algebras, and Noether's theorem.
In the second part, some applications of symmetries  
in physics are discussed, ranging from isospin and flavor symmetry 
to more recent developments involving the interacting boson model 
and its extension to supersymmetries in nuclear physics. 

\end{abstract}

\section{Introduction}

Symmetry and its mathematical framework---group theory---play an
increasingly important role in physics. Both classical and
quantum  systems usually display great complexity, but the
analysis of  their symmetry properties often gives rise to
simplifications and  new insights which can lead to a deeper
understanding. In  addition, symmetries themselves can point the
way toward the formulation of a correct physical theory by
providing constraints  and guidelines in an otherwise intractable
situation. It is  remarkable that, in spite of the wide variety of
systems one may  consider, all the way from classical ones to
molecules, nuclei, and elementary particles, group theory applies
the same basic  principles and extracts the same kind of useful
information from  all of them. This universality in the
applicability of symmetry considerations is one of the most
attractive features of group  theory.  Most people have an
intuitive understanding of symmetry,  particularly in its most
obvious manifestation in terms of geometric transformations that
leave a body or system invariant. This interpretation, however, is
not enough to readily grasp its deep connections with physics, and
it thus becomes necessary to generalize the notion of symmetry
transformations to encompass more abstract ideas. The mathematical
theory of these transformations is the subject matter of group
theory. 

Group theory was developed in the beginning of the 19th century by 
Evariste Galois (1811-1832) who pointed out the relation between the existence 
of algebraic solutions of a polynomial equation and the group of 
permutations associated with the equation. Another important contribution 
was made in the 1870's by Sophus Lie (1842-1899) 
who studied the mathematical theory of continuous transformations 
which led to the introduction of the basic concepts and operations 
of what are now known as Lie groups and Lie algebras. 
The deep connection between the abstract world of symmetries and 
dynamics---forces and motion and the fundamental laws of nature---was 
elucidated by Emmy Noether (1882-1935) 
in the early 20th century \cite{MacTutor}. 

The concept of symmetry has played a major role in physics, especially 
in the 20th century with the development of quantum mechanics and 
quantum field theory. There is an enormously wide range of applications 
of symmetries in physics. Some of the most important ones are listed below 
\cite{Iachello}.
\begin{itemize}
\item {\it Geometric symmetries} describe the arrangement of constituent 
particles into a geometric structure, for example the atoms in a molecule. 
\item {\it Permutation symmetries} in quantum mechanics lead to 
Fermi-Dirac and Bose-Einstein statistics for a system of identical 
particles with half-integer spin (fermions) and integer spin (bosons), 
respectively.
\item {\it Space-time symmetries} fix the form of the 
equations governing the motion of the constituent particles. For example, 
the form of the Dirac equation for a relativistic spin-1/2 particle
\begin{eqnarray*}
(i \gamma^{\mu} \partial_{\mu} - m ) \psi(x)=0 
\end{eqnarray*}
is determined by Lorentz invariance. 
\item {\it Gauge symmetries} fix the form of the interaction between 
constituent particles and external fields. For example, the form of the 
Dirac equation for a relativistic spin-1/2 particle in an external 
electromagnetic field $A_{\mu}$
\begin{eqnarray*}
\left[ \gamma^{\mu}(i \partial_{\mu} - e A_{\mu})-m\right] \psi(x) = 0 
\end{eqnarray*}
is dictated by the gauge symmetry of the electromagnetic interaction. 
The (electro-)weak and strong interactions are also governed by gauge symmetries. 
\item {\it Dynamical symmetries} fix the form of the interactions between 
constituent particles and/or external fields and determine the spectral 
properties of quantum systems. An early example was discussed by Pauli 
in 1926 \cite{Pauli} who recognized that the Hamiltonian of a particle in a 
Coulomb potential is invariant under four-dimensional rotations generated 
by the angular momentum and the Runge-Lenz vector.
\end{itemize}
 
These lectures notes are organized as follows. In the first part, a 
brief review is given of some of the basic mathematical concepts 
needed for the understanding of symmetries in nature, namely that  
of group theory, Lie groups and Lie algebras, and Noether's theorem.
In the second part, these ideas are illustrated by some applications 
in physics, ranging from isospin and flavor symmetry to more 
recent developments involving the interacting boson model and its extension 
to supersymmetries in nuclear physics. Some recent review articles on the 
concept of symmetries in physics are \cite{Iachello,Piet,oromana}.

\section{Elements of group theory}

In this section, some general properties of group theory are 
reviewed. For a more thorough discussion of the basic concepts and 
its properties, the reader is referred to the literature 
\cite{hamermesh,lipkin,gilmore,wybourne,elliott,georgi,stancu}.

\subsection{Definition of a group}

The concept of a group was introduced by Galois in a study of  
the existence of algebraic solutions of a polynomial equations. 
An abstract group $G$ is defined by a set of elements 
($\hat G_i, \hat G_j, \hat G_k, \ldots$) 
for which a ``multiplication'' rule (indicated here by $\circ$) combining 
these elements exists and which satisfies the following conditions. 
\begin{itemize}
\item {\it Closure.} \\
If $\hat G_i$ and $\hat G_j$
are elements of the set, so is their product $\hat G_i \circ \hat G_j ~.$
\item {\it Associativity.} \\
The following property is always valid:
$$\hat G_i \circ (\hat G_j \circ \hat G_k)=(\hat G_i \circ \hat G_j) \circ \hat G_k ~.$$
\item {\it Identity.} \\
There exists an element $\hat E$ of $G$ satisfying
$$\hat E \circ \hat G_i=\hat G_i \circ \hat E=\hat G_i ~.$$
\item {\it Inverse.} \\
For every $\hat G_i$
there exists an element  $\hat G_i^{-1}$ such that
$$\hat G_i \circ \hat G_i^{-1}=\hat G_i^{-1} \circ \hat G_i=\hat E ~.$$
\end{itemize}
The number of elements is called the {\it order} of the group.
If in addition the elements of a group satisfy the condition 
of commutativity, the group is called an Abelian group. 
\begin{itemize}
\item {\it Commutativity.} \\
All elements obey $$\hat G_i \circ \hat G_j = \hat G_j \circ \hat G_i ~.$$
\end{itemize}

\subsection{Lie groups and Lie algebras}

For continuous (or Lie) groups all elements may be
obtained by exponentiation in terms of a basic set of elements
$\hat g_i$, $i=1,2,\dots,s$, called {\it generators}, which together
form the {\it Lie algebra} associated with the Lie group. A simple
example is provided by the group of rotations in
two-dimensional space, with elements that may be realized as
\begin{equation}
\hat G(\alpha)=\exp[-i\alpha\hat l_z] ~, 
\label{ad1}
\end{equation}
where $\alpha$ is  the angle of rotation and
\begin{equation}
\hat l_z=-i\left(x{\frac{\partial}{\partial y}}
-y{\frac{\partial}{\partial x}} \right) ~,
\label{ad2}
\end{equation}
is the generator of these
transformations in the $x$--$y$ plane. Three-dimensional
rotations  require the introduction of two additional generators,
associated with rotations in the $z$--$x$ and $y$--$z$ planes,
\begin{equation}
 \hat l_y=-i\left(z{\frac{\partial}{\partial x}}
-x{\frac{\partial}{\partial z}}\right)~, \qquad
 \hat l_x=-i\left(y{\frac{\partial}{\partial z}}
-z{\frac{\partial}{\partial y}}\right) ~,
\label{ad3}
\end{equation}
Finite rotations can then be parametrized by
three angles (which may be chosen to be the Euler angles) and
expressed as a product of exponentials of the generators
of Eqs.~(\ref{ad2}) and (\ref{ad3}). Evaluating the
commutators of these operators, we find
\begin{equation} [\hat l_x,\hat l_y]=i\hat l_z~,
\qquad [\hat l_y,\hat l_z]=i\hat l_x~, 
\qquad [\hat l_z,\hat l_x]=i\hat l_y~,
\label{ad4}
\end{equation}
which illustrates the closure property of the
group generators. In general, the operators
$\hat g_i$, $i=1,2,\dots,s$, define a {\it Lie algebra} if they close
under commutation 
\begin{equation} [\hat g_i,\hat g_j]=\sum_k c^k_{ij}\hat g_k ~,
\label{ad5}
\end{equation}
and satisfy the Jacobi identity 
\begin{equation}
[\hat g_i,[\hat g_j,\hat g_k]]+[\hat g_k,[\hat g_i,\hat g_j]]
+[\hat g_j,[\hat g_k,\hat g_i]]=0~.
\label{ad6}
\end{equation}
The constants $c^k_{ij}$ are called {\it structure constants}, 
and determine the properties of both the Lie algebra and its 
associated Lie group. Lie groups have been 
classified by Cartan, and many of their properties have
been established. 

The group of unitary transformations in $n$ dimensions is
denoted by $U(n)$ and of rotations in $n$ dimensions by $SO(n)$ 
(Special Orthogonal). The corresponding Lie algebras are sometimes 
indicated by lower case symbols, $u(n)$ and $so(n)$, respectively. 

\subsection{Symmetries and conservation laws}

Symmetry in physics is expressed by the invariance of a 
Lagrangian or of a Hamiltonian or, equivalently, of the 
equations of motion, with respect to some group of transformations. 
The connection between the abstract concept of symmetries 
and dynamics is formulated as Noether's theorem which says that, 
irrespective of a classical or a quantum mechanical treatment, 
an invariant Lagrangian or Hamiltonian with respect to a 
continuous symmetry implies a set of conservation laws \cite{hill}. 
For example, the conservation of energy, momentum and angular 
momentum are a consequence of the invariance of the system 
under time translations, space translations and rotations, 
respectively.  

In quantum mechanics, continuous symmetry transformations can 
in general be expressed as 
\begin{equation}
U=\exp\left[ i \sum_j \alpha_j \hat{g}_j \right] ~.
\end{equation}
States and operators transform as 
\begin{equation}
\left| \psi \right> \rightarrow \left| \psi' \right> = U \left| \psi \right> ~, 
\hspace{1cm} A \rightarrow A'=UAU^{\dagger} ~.
\end{equation}
For the Hamiltonian one then has 
\begin{equation}
H \rightarrow H'=UHU^{\dagger}=H + i \sum_j \alpha_j [\hat{g}_j,H]+ {\cal O}(\alpha^2) ~.
\end{equation}
When the physical system is invariant under the symmetry transformations $U$,  
the Hamiltonian remains the same $H'=H$. Therefore, the Hamiltonian 
commutes with the generators of the symmetry transformation
\begin{equation}
\left[ \hat{g}_j,H \right]=0 ~,
\label{at9}
\end{equation}
which implies that the generators are constants of the motion. 
Eq.~(\ref{at9}), together with the closure relation of the generators 
of Eq.~(\ref{ad5}), constitutes the definition of the symmetry algebra 
for a time-independent system.

\subsection{Constants of the motion and state labeling}
\label{constants}

For any Lie algebra one may construct one or more operators 
$\hat {\cal C}_l$ which commute with all the generators $\hat g_j$ 
\begin{equation}
[\hat {\cal C}_l,\hat g_j]=0, \qquad l=1,2,\dots,r, \quad j=1,2,\dots,s~.
 \end{equation}
These operators are called {\it Casimir operators} or 
{\it Casimir invariants}. They may be linear, quadratic, or higher
 order in the generators. The number $r$ of linearly independent
 Casimir operators is called the rank of the algebra \cite{wybourne}. 
 This number coincides with the maximum subset of
 generators which commute among themselves (called {\it weight
 generators})
\begin{equation}
[\hat g_\alpha,\hat g_\beta]=0~, \qquad \alpha,\beta=1,2,\dots,r~,
\label{ac2}
\end{equation}
where greek labels were used to indicate that they belong to the subset 
satisfying Eq.~(\ref{ac2}). The operators $(\hat{\cal C}_l$, $\hat g_\alpha)$ 
may be simultaneously diagonalized and their eigenvalues used to label
 the corresponding eigenstates.

To illustrate these definitions, we consider the $su(2)$ algebra
$(\hat j_x$, $\hat j_y$, $\hat j_z)$ with commutation relations
\begin{equation}
 [\hat j_x,\hat j_y]=i\hat j_z~, \qquad [\hat j_z,\hat j_x]=i\hat j_y~, \qquad
 [\hat j_y,\hat j_z]=i\hat j_x~, \label{ac3}
\end{equation}
which is isomorphic to the $so(3)$ commutators given in Eq.~(\ref{ad4}).
 From Eq.~(\ref{ac3}) one can conclude that the rank of the algebra is $r=1$. 
Therefore one can choose $\hat j_z$ as the generator to diagonalize together
 with the Casimir invariant
\begin{equation}
\hat j^2=\hat j^2_x+\hat j^2_y+\hat j^2_z~.
 \end{equation}
The eigenvalues and branching rules for the commuting set
 $(\hat {\cal C}_l$, $\hat g_\alpha)$ can be determined solely from the commutation
 relations Eq.~(\ref{ad5}). In the case of $su(2)$ the eigenvalue
 equations are
\begin{equation}
\hat j^2\vert jm\rangle=n_j\vert jm\rangle~,
 \qquad \hat j_z\vert jm\rangle=m\vert jm\rangle~,
 \end{equation}
where $j$ is an index to distinguish the different $\hat j^2$
 eigenvalues. Defining the raising and lowering operators
\begin{equation}
 \hat j_\pm=\hat j_x\pm i\hat j_y~,
 \end{equation}
and using Eq.~(\ref{ac3}), one finds the well-known results 
\begin{equation}
n_j=j(j+1)~, \qquad j=0,\frac{1}{2},1,\ldots,
 \qquad m=-j,-j+1,\ldots,j~.
 \end{equation}
As a bonus, the action of $\hat j_\pm$ on the $\vert jm\rangle$
 eigenstates is determined to be
\begin{equation}
\hat j_\pm \vert jm\rangle = \sqrt{(j\mp m)(j\pm m+1)} \vert jm\pm 1 \rangle~.
 \end{equation}
In the case of a general Lie algebra, see Eq.~(\ref{ad5}), this 
 procedure becomes quite complicated, but it requires the same basic
 steps. The analysis leads to the algebraic determination of
 eigenvalues, branching rules, and matrix elements of raising and
 lowering operators \cite{wybourne}.
 
The {\em symmetry} algebra
 provides constants of the motion, which in turn lead to quantum
 numbers that label the states associated with a given energy
 eigenvalue. The raising and lowering operators in this algebra
 only connect degenerate states. The dynamical algebra, however,
 defines the whole set of eigenstates associated with a given
 system. The generators are no longer constants of the motion as
 not all commute with the Hamiltonian. The raising and lowering
 operators may now connect all states with each other.

\subsection{Dynamical symmetries}
\label{algebraic}

In this section we
show how the concepts presented in the previous sections lead to
an algebraic approach which can be applied to the study of different 
physical systems.
 We start by considering again Eq.~(\ref{at9}) which
describes the invariance of a Hamiltonian under the algebra
$g \equiv (\hat g_j)$
\begin{equation} [H,\hat g_j]=0~,
\label{aa1}
\end{equation}
implying that $g$
plays the role of symmetry algebra for the system. An eigenstate of
$H$ with energy $E$ may be written as $\vert \Gamma \gamma \rangle$,
where $\Gamma$ labels the irreducible representations of the
group $G$ corresponding to $g$ and $\gamma$ distinguishes between the
different eigenstates with energy $E$ (and may be chosen to
correspond to irreducible representations of subgroups of $G$).
The energy eigenvalues of the Hamiltonian in Eq.~(\ref{aa1}) thus
depend only on $\Gamma$ 
\begin{equation}
H \vert \Gamma \gamma \rangle
=E(\Gamma) \vert \Gamma \gamma \rangle~.
 \end{equation}
The generators $\hat g_j$ do not admix states with different $\Gamma$'s.
 
Let's now consider the chain of algebras
\begin{equation} g_1\supset g_2~,
 \end{equation}
which will lead to the introduction of the concept of {\it dynamical symmetry}. 
Here $g_2$ is a subalgebra of $g_1$, $g_2 \subset g_1$, {\it i.e.} its generators 
form a subset of the generators of $g_1$ and close under commutation.  
 If $g_1$ is a symmetry algebra for $H$, its 
 eigenstates can be labeled as $\vert \Gamma_1 \gamma_1 \rangle$. Since
 $g_2 \subset g_1$, $g_2$ must also be a symmetry algebra for $H$
 and, consequently, its eigenstates labeled as
 $\vert \Gamma_2 \gamma_2\rangle$.  Combination of the two properties
 leads to the eigenequation
\begin{equation}
H \vert \Gamma_1 \Gamma_2 \gamma_2\rangle
 =E(\Gamma_1) \vert \Gamma_1 \Gamma_2 \gamma_2\rangle~,
\label{aa4}
\end{equation}
 where the role of $\gamma_1$ is played by $\Gamma_2 \gamma_2$ and hence
 the eigenvalues depend only on $\Gamma_1$. This process may be
 continued when there are further subalgebras, that is, $g_1\supset
 g_2\supset g_3\supset\cdots$, in which case $\gamma_2$ is substituted
 by $\Gamma_3 \gamma_3$, and so on.

 In many physical applications the original assumption that $g_1$
 is a symmetry algebra of the Hamiltonian is found to be too strong
 and must be relaxed, that is, one is led to consider the breaking
 of this symmetry. An elegant way to do so is by considering a
 Hamiltonian of the form
\begin{equation} 
H'=a \, \hat {\cal C}_{l_1}(g_1)+b \, \hat{\cal C}_{l_2}(g_2)~,
 \label{aa5}
\end{equation}
where $\hat{\cal C}_{l_i}(g_i)$ is a Casimir invariant of
 $g_i$. Since $[H',\hat g_i]=0$ for $\hat g_i\in g_2$, $H'$ is
 invariant under $g_2$, but not anymore under $g_1$ because
 $[\hat{\cal C}_{l_2}(g_2),\hat g_i]\neq 0$ for $\hat g_i\not\in g_2$.
The new {\it symmetry algebra} is thus $g_2$ while $g_1$ now plays the role of
 {\it dynamical algebra} for the system, as long as all states we
 wish to describe are those originally associated with
 $E(\Gamma_1)$. The extent of the symmetry breaking depends on the
 ratio $b/a$. Furthermore, since $H'$ is given as a combination
 of Casimir operators, its eigenvalues can be obtained in closed form
\begin{equation}
H' \vert \Gamma_1 \Gamma_2 \gamma_2\rangle
 = \left[ a \, E_{l_1}(\Gamma_1)+b \, E_{l_2}(\Gamma_2) \right]
 \vert \Gamma_1 \Gamma_2 \gamma_2\rangle~.
\label{aa6}
\end{equation}
The kind of
 symmetry breaking caused by interactions of the form (\ref{aa5})
 is known as {\it dynamical-symmetry breaking} and the remaining
 symmetry is called a {\it dynamical symmetry} of the Hamiltonian
 $H'$.
 From Eq.~(\ref{aa6}) one concludes that
 even if $H'$ is not invariant under $g_1$, its eigenstates are
 the same as those of $\hat H$ in Eq.~(\ref{aa4}). The dynamical-symmetry
 breaking thus splits but does not admix the eigenstates.

In the last part of these lecture notes, we discuss some 
applications of the algebraic approach in nuclear and particle 
physics. The algebraic approach, both in the sense we have defined 
here and in its generalizations to other fields of research, has become an
important tool in the search for a unified description of physical
phenomena. 

\section{Isospin symmetry}
 
\begin{figure}[t]
\centerline{\epsfig{file=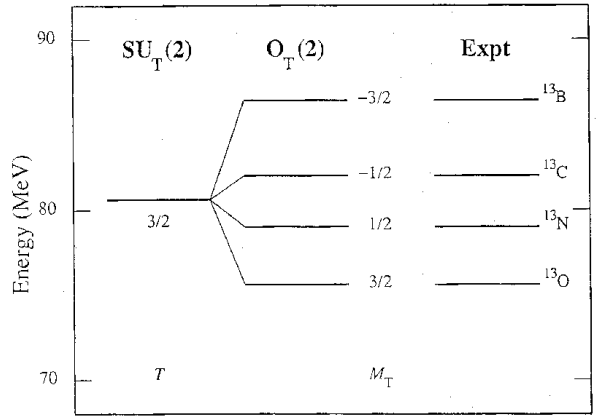,width=0.5\textwidth}} 
\caption[]{\small
Binding energies of the $T=3/2$ isobaric analog states with
angular momentum and parity $J^\pi=1/2^-$ in $^{13}$B,
$^{13}$C, $^{13}$N, and $^{13}$O. The column on the left is
obtained for an exact $SU_T(2)$ {\em symmetry}, which predicts
states with different $M_T$ to be degenerate. The middle column is
obtained in the case of an $SU_T(2)$ {\em dynamical symmetry},
 Eq.~(\ref{aa7}) with parameters $a=80.59$, $b=-2.96$, and
 $c=-0.26$ MeV.}
\label{isobaric}
\end{figure}

 Some of these ideas can be illustrated with well-known examples.
 In 1932 Heisenberg considered the occurrence of isospin multiplets
 in nuclei \cite{aHEI32}. To a good approximation, the strong interaction 
 between nucleons does not distinguish between protons and neutrons. 
 In the isospin formalism, the proton and neutron are treated as one 
 and the same particle: the nucleon with isospin $t=\frac{1}{2}$. 
 The isospin projections $m_t=+\frac{1}{2}$ and $-\frac{1}{2}$ are 
 identified with the proton and the neutron, respectively. The total 
 isospin of the nucleus is denoted by $T$ and its projection by $M_T$.  
 In the notation
used above (without making the distinction between algebras and
 groups), $G_1$ is in this case the isospin group $SU_T(2)$ 
 generated by the operators $\hat T_x$, $\hat T_y$,
and $\hat T_z$ which
 satisfy commutation relations of Eq.~(\ref{ac3}),
and $G_2$ can be identified with $SO_T(2)$ generated by $\hat T_z)$. An
isospin-invariant Hamiltonian commutes with $\hat T_x$, $\hat T_y$, and
$\hat T_z$, and hence the eigenstates $\vert T M_T \rangle$ with fixed
$T$ and $M_T=-T,-T+1,\ldots,T$ are degenerate in energy. 
However, the electromagnetic interaction breaks isospin invariance 
due to difference in electric charge of the proton and the neutron, and 
lifts the degeneracy of the states $\vert T M_T\rangle$. It is assumed that
this symmetry breaking occurs dynamically, and since the Coulomb
force has a two-body character, the breaking terms are at most
quadratic in $\hat T_z$ \cite{elliott}. The energies of the
corresponding nuclear states with the same $T$ are then given by
\begin{equation} 
E(M_T)=a+b \, M_T + c \, M_T^2~,
\label{aa7}
\end{equation}
and $SU_T(2)$
becomes the dynamical symmetry for the system while $SO_T(2)$ is
the symmetry algebra. The dynamical symmetry breaking thus implied
that the eigenstates of the nuclear Hamiltonian have well-defined
values of $T$ and $M_T$. Extensive tests have shown that indeed
this is the case to a good approximation, at least at low
excitation energies and in light nuclei \cite{aBOH75}. 
Eq.~(\ref{aa7}) can be tested in a number of cases. In
Figure~\ref{isobaric} a
$T=3/2$ multiplet consisting of states in the nuclei $^{13}$B,
$^{13}$C, $^{13}$N, and $^{13}$O is compared with the theoretical
 prediction of Eq.~(\ref{aa7}).

\section{Flavor symmetry}

 A less trivial example of dynamical-symmetry breaking is provided
 by the Gell-Mann--Okubo mass-splitting formula for elementary
 particles \cite{aGEL62,aOKU62}. In the previous example, we saw 
 that the near equality of the neutron and proton masses suggested 
 the existence of isospin multiplets which was later confirmed at 
 higher energies for other particles. Gell-Mann and Ne'eman proposed 
 independently a dynamical algebra to further classify and order 
 these different isospin 
 multiplets of hadrons in terms of $SU(3)$ representations \cite{aGEL64} . 
 Baryons were found to occur in decuplets, octets and singlets, whereas 
 mesons appear only in octets and singlets. The members of a $SU(3)$ 
 multiplet are labeled by their isospin $T$, $M_T$ and hypercharge $Y$ 
 quantum numbers, according to the group chain 
\begin{equation}
 \begin{array}{ccccccccc}
 SU(3) &\supset& U_Y(1) &\otimes& SU_T(2) &\supset& U_Y(1) &\otimes& SO_T(2) \\
 \downarrow && \downarrow && \downarrow &&&& \downarrow \\
 (\lambda,\mu) && Y && T &&&& M_T \end{array}
 \label{aa8}
\end{equation}
If one would assume $SU(3)$ invariance, all
 particles in a multiplet would have the same mass, but since the
 experimental masses of other baryons differ from the nucleon
 masses by hundreds of MeV, the $SU(3)$ symmetry clearly must be
 broken.

\begin{figure}[t]
\centerline{\epsfig{file=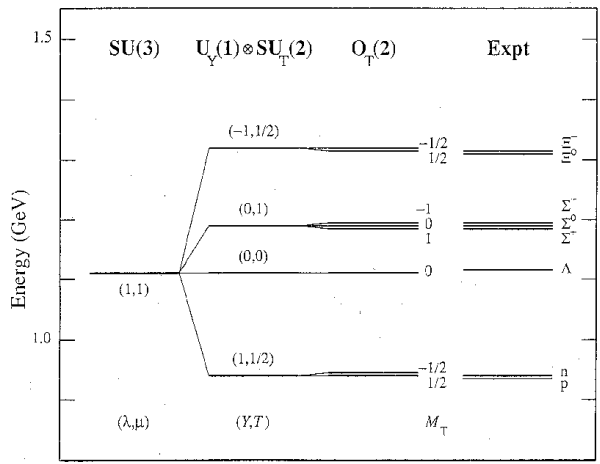,width=0.5\textwidth}} 
\caption{\small Mass spectrum of the ground state baryon octet. 
The column on the left is obtained for an exact $SU(3)$ symmetry, 
which predicts all masses to be the same, while the next two columns 
represent successive breakings of this symmetry in a dynamical manner. 
The column under $SO_T(2)$ is obtained with Eq.~(\ref{okubo}) with 
parameters $a=1111.3$, $b=-189.6$, $d=-39.9$, $e=-3.8$, and $f=0.9$ MeV.}
\label{octete}
\end{figure}

 Dynamical symmetry breaking allows the baryon states to still be
 classified by Eq.~(\ref{aa8}). Following the procedure outlined above
 and keeping up to quadratic terms, one finds a mass operator of
 the form
\begin{eqnarray}
 \hat M&=&a+b \, \hat{\cal C}_{1U_Y(1)}+c \, \hat{\cal C}^2_{1U_Y(1)}
+d \, \hat{\cal C}_{2SU_T(2)} +e \, \hat{\cal C}_{1SO_T(2)} 
+f \, \hat{\cal C}^2_{1SO_T(2)} ~,
\end{eqnarray}
with eigenvalues
\begin{equation}
M(Y,T,M_T)=a+b\,Y+c\,Y^2+d\,T(T+1)+e\,M_T+f\,M_T^2~.
\label{aa10}
\end{equation}
A further assumption regarding the $SU(3)$ tensor character of the 
strong interaction \cite{aGEL62,aOKU62} leads to a relation between 
the coefficients $c$ and $d$ in Eq.~(\ref{aa10}), $c=-d/4$
\begin{equation}
M'(Y,T)=a+b \, Y+d\left[T(T+1)-\frac{1}{4}Y^2 \right]+e\,M_T+f\,M_T^2~.
\label{okubo}
\end{equation}
If one neglects the isospin breaking due to the last two terms, 
one recovers the Gell-Mann-Okubo mass formula. 
In Figure~\ref{octete} this process of successive dynamical-symmetry
 breaking is illustrated with the octet representation containing
 the neutron and the proton and the $\Lambda$, $\Sigma$, and $\Xi$
 baryons. 
 
\def\bx{{\bar x}}

\def\bfD{{\bf D}}
\def\bfG{{\bf G}}
\def\bfM{{\bf M}}
\def\bfR{{\bf R}}
\def\bfS{{\bf S}}
\def\bfT{{\bf T}}
\def\bfU{{\bf U}}


\def\bfmh{{\vec{h}}}
\def\bfml{{\bf l}}
\def\bfmL{{\bf L}}
\def\bfmp{{\bf p}}
\def\bfmr{{\bf r}}
\def\bfmR{{\bf R}}
\def\bfmx{{\bf x}}
\def\bfom{{\vec{\omega}}}
\def\bfps{{\vec{\psi}}}

\def\bfoms{{\vec{\omega}}}

\def\cA{{\cal A}}
\def\cB{{\cal B}}
\def\cC{{\cal C}}
\def\cD{{\cal D}}
\def\cG{{\cal G}}
\def\cN{{\cal N}}
\def\cO{{\cal O}}
\def\cP{{\cal P}}
\def\cdP{{\cal P}^\dagger}
\def\ctG{\tilde{\cal G}}
\def\ctP{\tilde{\cal P}}
\def\cQ{{\cal Q}}
\def\cY{{\cal Y}}

\def\da{a^\dagger}
\def\db{b^\dagger}
\def\dd{d^\dagger}
\def\dh{h^\dagger}
\def\dpz{p^\dagger}
\def\ds{s^\dagger}
\def\dt{t^\dagger}

\def\hA{{\hat A}}
\def\hB{{\hat B}}
\def\hD{{\hat D}}
\def\he{{\hat e}}
\def\hE{{\hat E}}
\def\hF{{\hat F}}
\def\hg{{\hat g}}
\def\hG{{\hat G}}
\def\hH{{\hat H}}
\def\hj{{\hat\jmath}}
\def\hJ{{\hat J}}
\def\hk{{\hat k}}
\def\hK{{\hat K}}
\def\hl{{\hat l}}
\def\hL{{\hat L}}
\def\hm{{\hat m}}
\def\hM{{\hat M}}
\def\hn{{\hat n}}
\def\hN{{\hat N}}
\def\hO{{\hat O}}
\def\hP{{\hat P}}
\def\hQ{{\hat Q}}
\def\hR{{\hat R}}
\def\hS{{\hat S}}
\def\hT{{\hat T}}
\def\hU{{\hat U}}
\def\hV{{\hat V}}
\def\hX{{\hat X}}

\def\hcN{{\hat{\cal N}}}

\def\oh{\bar{h}}
\def\oO{\overline{\rm O}}
\def\osigma{\bar{\sigma}}
\def\oSO{\overline{\rm SO}}
\def\oSU{\overline{\rm SU}}
\def\oU{\overline{\rm U}}

\def\rb{{\rm b}}
\def\rbent{{\rm bent}}
\def\rB{{\rm B}}
\def\rBF{{\rm BF}}
\def\rd{{\rm d}}
\def\re{{\rm e}}
\def\rE{{\rm E}}
\def\reven{{\rm even}}
\def\rf{{\rm f}}
\def\rF{{\rm F}}
\def\rg{{\rm g}}
\def\rhw{{\rm hw}}
\def\ri{{\rm i}}
\def\rI{{\rm I}}
\def\rII{{\rm II}}
\def\rIII{{\rm III}}
\def\rJ{{\rm J}}
\def\rlin{{\rm lin}}
\def\rM{{\rm M}}
\def\rN{{\rm N}}
\def\rns{{\rm ns}}
\def\ro{{\rm o}}
\def\rO{{\rm O}}
\def\rodd{{\rm odd}}
\def\rp{{\rm p}}
\def\rr{{\rm r}}
\def\rR{{\rm R}}
\def\rrv{{\rm rv}}
\def\rrig{{\rm rig}}
\def\rrve{{\rm rv-e}}
\def\rs{{\rm s}}
\def\rsd{{\rm sd}}
\def\rSO{{\rm SO}}
\def\rSp{{\rm Sp}}
\def\rst{{\rm st}}
\def\rSU{{\rm SU}}
\def\rt{{\rm t}}
\def\ru{{\rm u}}
\def\rU{{\rm U}}
\def\rv{{\rm v}}
\def\rx{{\rm x}}

\def\ta{{\tilde a}}
\def\tb{{\tilde b}}
\def\td{{\tilde d}}
\def\tep{{\tilde\epsilon}}
\def\tK{{\tilde K}}
\def\tl{{\tilde l}}
\def\tn{{\tilde n}}
\def\tp{{\tilde p}}
\def\ts{{\tilde s}}
\def\tSO{{\widetilde\rSO}}
\def\tt{{\tilde t}}

\def\seq{\!=\!}
\def\sitem{\item\vspace{-3mm}}
\def\smin{\!-\!}
\def\smp{\!\mp\!}
\def\sparallel{\!\parallel\!}
\def\splus{\!+\!}
\def\spm{\!\pm\!}
\def\ssmin{\!\!-\!\!}
\def\ssplus{\!\!+\!\!}
\def\stimes{\!\times\!}

\def\mink{\!-\!}
\def\mpk{\!\mp\!}
\def\plusk{\!+\!}
\def\pmk{\!\pm\!}

\def\dscal{\!:\!}
\def\scal{\cdot}

\def\tc{\!\times\!}

\def\tf#1#2{
\raise0.40ex\hbox{$\scriptstyle#1$}
\raise0.20ex\hbox{$\scriptstyle/$}
\raise0.00ex\hbox{$\scriptstyle#2$}}
\def\to#1#2{{\textstyle{#1\over#2}}}
\def\spin#1{
\raise0.40ex\hbox{$\scriptstyle#1$}
\raise0.20ex\hbox{$\scriptstyle/$}
\raise0.00ex\hbox{$\scriptstyle2$}}

\def\pder#1{{{\partial}\over{\partial#1}}}
\def\pderm#1#2{{{\partial^{#1}}\over{\partial{#2}^{#1}}}}
\def\der#1{{{d}\over{d#1}}}
\def\derm#1#2{{{d^{#1}}\over{d{#2}^{#1}}}}

\def\bin#1#2{\biggl(\!\begin{array}{c}#1\\#2\end{array}\!\biggr)}
\def\pthree#1#2#3#4#5#6{\biggl(\!\begin{array}{ccc}
#1&#2&#3\\#4&#5&#6
\end{array}\!\biggr)}
\def\racah#1#2#3#4#5#6{\biggl\{\!\begin{array}{ccc}
#1&#2&#3\\#4&#5&#6
\end{array}\!\biggr\}}
\def\ninej#1#2#3#4#5#6#7#8#9{\left\{\!\begin{array}{ccc}
#1&#2&#3\\#4&#5&#6\\#7&#8&#9
\end{array}\!\right\}}
\def\xninej#1#2#3#4#5#6#7#8#9{\left[\!\begin{array}{ccc}
#1&#2&#3\\#4&#5&#6\\#7&#8&#9
\end{array}\!\right]}
\def\coupt#1#2#3#4#5#6{\langle#1#2\otimes#3#4\vert#5#6\rangle}
\def\coup#1#2#3#4#5#6{\left\langle\!\begin{array}{cc|c}
#1&#3&#5\\#2&#4&#6
\end{array}\!\right\rangle}

\def\tauv{{v}}

\def\dsum#1#2{{\displaystyle \sum_{{\scriptstyle #1}\atop{\scriptstyle #2}}}}
\def\sumd#1{{\displaystyle \sum_{#1}}}

\def\youngonerow#1{\begin{array}{l}
\overbrace{\fox\fox\cdots\fox}^{#1}
\end{array}}
\def\youngonerowone#1{\begin{array}{l}
\overbrace{\fox\fox\fox\cdots\fox}^{#1}\\
\fox
\end{array}}
\def\youngtworow#1#2{\begin{array}{l}
\overbrace{\fox\fox\fox\cdots\fox}^{#1}\\
\overbrace{\fox\fox\cdots\fox}^{#2}
\end{array}}
\def\youngnrow#1#2#3{\begin{array}{l}
\overbrace{\fox\fox\fox\cdots\fox}^{#1}\\
\overbrace{\fox\fox\cdots\fox}^{#2}\\
\vdots\\
\overbrace{\fox\cdots\fox}^{#3}
\end{array}}
\def\youngonecolumn#1{\left.\begin{array}{c}
\fox\\\fox\\\vdots\\\fox
\end{array}
\right\}#1}
\def\youngtwocolumn#1#2{\left.\begin{array}{c}
\fox\\\fox\\\fox\\\vdots\\\fox
\end{array}
\right\}#1
\left.\begin{array}{c}
\fox\\\fox\\\vdots\\\fox
\end{array}
\right\}#2}

\def\youngone{\begin{array}{c}
\fox
\end{array}}
\def\youngoneone{\begin{array}{c}
\fox\\\fox
\end{array}}
\def\youngoneoneone{\begin{array}{l}
\fox\\\fox\\\fox
\end{array}}
\def\youngoneoneoneone{\begin{array}{l}
\fox\\\fox\\\fox\\\fox
\end{array}}
\def\youngtwo{\fox\fox}
\def\youngtwooneone{\begin{array}{l}
\fox\fox\\\fox\\\fox
\end{array}}
\def\youngtwotwo{\begin{array}{l}
\fox\fox\\\fox\fox
\end{array}}
\def\youngtwotwooneone{\begin{array}{l}
\fox\fox\\\fox\fox\\\fox\\\fox
\end{array}}
\def\youngtwotwotwo{\begin{array}{l}
\fox\fox\\\fox\fox\\\fox\fox
\end{array}}

\def\fox{\mbox{\large$\Box$}}
\def\foxa{\fox\kern-0.65em\raise0.55ex\hbox{$\scriptstyle{a}$}\kern0.27em}
\def\foxb{\fox\kern-0.63em\raise0.40ex\hbox{$\scriptstyle{b}$}\kern0.25em}

\def\chainspace{\arraycolsep=0.15em}

\def\References{
\section*{References}
\markright{\bf References}}
\def\Summary{
\section*{Summary}
\markright{\bf Summary}}

\def\refboo#1#2#3#4#5
{#1, {\it#2}, #3, #4, #5}
\def\refart#1#2#3#4#5#6
{#1, ``#2,'' #3 {\bf#4} (#5) #6}
\def\refarts#1#2#3#4#5
{#2 {\bf#3} (#4) #5}
\def\refcon#1#2#3#4#5#6#7#8
{#1, ``#2,'' in {\it#3}, edited by #4, #5, #6, #7, p.~#8}
\def\refthe#1#2#3#4
{#1, {\it#2}, doctoral dissertation, #3, #4}

\def\beq{\begin{equation}}
\def\eeq{\end{equation}}
\def\beqn{\begin{equation}} 
\def\eeqn{\end{equation}} 
\def\beqa{\begin{eqnarray}}
\def\eeqa{\end{eqnarray}}
\def\beqan{\begin{eqnarray}} 
\def\eeqan{\end{eqnarray}} 
\def\non{\nonumber\\}

\def\dim{{\rm dim}}

\hyphenation{
ana-logy
di-men-sion di-men-sions di-men-sion-al
ei-gen-func-tion ei-gen-func-tions
ei-gen-state ei-gen-states
ei-gen-val-ue ei-gen-val-ues
in-fini-tesi-mal
mol-ecule mol-ecules mol-ecu-lar
quad-rat-ic
ro-ta-tion ro-ta-tions ro-ta-tion-al
single
sym-me-try sym-me-tries sym-met-ric anti-sym-met-ric
su-per-sym-me-try su-per-sym-me-tries su-per-sym-met-ric
vi-bra-tion vi-bra-tions vi-bra-tion-al
wheth-er}

\section{Nuclear supersymmetry}

Nuclear supersymmetry (n-SUSY) is a composite-particle phenomenon,
linking the properties of bosonic and fermionic systems, framed in
the context of the Interacting Boson Model of nuclear structure \cite{IBM}. 
Composite particles, such as the $\alpha$-particle are  known to
behave as approximate bosons.  As He atoms they become superfluid
at low temperatures, an under certain conditions can  also form
Bose-Einstein condensates.  At higher densities (or temperatures)
the constituent fermions begin to be felt and the  Pauli principle
sets in. Odd-particle composite systems, on the  other hand,
behave as approximate fermions, which in the case of the Interacting
Boson-Fermion Model are treated as a combination of bosons and an 
(ideal) fermion \cite{IBFM}. In
contrast to the theoretical construct of supersymmetric particle
physics, where SUSY is postulated as a generalization of the
Lorentz-Poincare invariance at a fundamental level, experimental
evidence has been found for n-SUSY 
\cite{FI,susy,baha,thesis,tres,pt195,au196} 
as we shall discuss below. 
Nuclear supersymmetry should not be confused with fundamental SUSY, which
predicts the existence of supersymmetric particles, such as the
photino and the selectron for which, up to now, no evidence
has been found. If such particles exist, however, SUSY must be
strongly broken, since large mass differences must exist among
superpartners, or otherwise they would have been already detected.
Nuclear supersymmetry, on the other hand,
is a theory that establishes precise links among the spectroscopic
properties of certain neighboring nuclei. Even-even and odd-odd
nuclei are composite bosonic systems, while odd-$A$ nuclei are
fermionic. It is in this context that n-SUSY provides a
theoretical framework where bosonic and fermionic systems
are treated as members of the same supermultiplet \cite{baha}.
Nuclear supersymmetry treats the excitation spectra and
transition intensities of the  different nuclei as arising from a
single Hamiltonian and a single set of transition operators. 
Nuclear supersymmetry was originally postulated as a
symmetry among pairs of nuclei \cite{FI,susy,baha}, and was
subsequently extended to quartets of nuclei,
where odd-odd nuclei could be incorporated in a natural
way \cite{quartet}.  Evidence for the existence of n-SUSY (albeit
possibly significantly broken) grew over the years, specially for
the quartet of nuclei $^{194}$Pt, $^{195}$Au, $^{195}$Pt
and $^{196}$Au, but only recently more systematic evidence
was found \cite{tres,pt195,au196}.

We first present a pedagogic review of dynamical (super)symmetries in 
even- and odd-mass nuclei, which is based in part on \cite{thesis}. 
Next we discuss the generalization of these concepts to include the 
neutron-proton degree of freedom. 

\subsection{Dynamical symmetries in even-even nuclei}

Dynamical supersymmetries were introduced \cite{FI} in nuclear physics in
1980 by Franco Iachello in the context of the Interacting Boson Model (IBM)
and its extensions. The spectroscopy of atomic nuclei is
characterized by the interplay between collective (bosonic) and
single-particle (fermionic) degrees of freedom.

The IBM describes collective excitations in even-even nuclei in
terms of a system of interacting monopole and quadrupole bosons with angular
momentum $l=0,2$ \cite{IBM}. The bosons are associated with the number of
correlated proton and neutron pairs, and hence the number of bosons $N$ is
half the number of valence nucleons. Since it is convenient to express
the Hamiltonian and other operators of interest in second quantized form,
we introduce creation, $s^{\dagger}$ and $d^{\dagger}_m$, and annihilation,
$s$ and $d_m$, operators, which altogether can be denoted by
$b^{\dagger}_{i}$ and $b_{i}$ with $i=l,m$ ($l=0,2$ and $-l \leq m \leq l$).
The operators $b^{\dagger}_{i}$ and $b_{i}$ satisfy the commutation
relations
\begin{equation}
[b_i,b^{\dagger}_j] \;=\; \delta_{ij} ~,
\hspace{1cm} [b^{\dagger}_i,b^{\dagger}_j]
\;=\; [b_i,b_j] \;=\; 0 ~.
\end{equation}
The bilinear products
\begin{equation}
B_{ij} \;=\; b^{\dagger}_i b_j ~,
\label{bosgen}
\end{equation}
generate the algebra of $U(6)$ the unitary group in 6 dimensions
\begin{equation}
[ B_{ij},B_{kl} ] \;=\; B_{il} \, \delta_{jk} - B_{kj} \, \delta_{il} ~.
\end{equation}
We want to construct states and operators that transform according
to irreducible representations of the rotation group (since the
problem is rotationally invariant).
The creation operators $b^{\dagger}_i$ transform by definition as
irreducible tensors under rotation. However, the annihilation operators
$b_i$ do not. It is an easy exercise to contruct operators that do
transform appropriately
\begin{equation}
\tilde{b}_{lm} \;=\; (-)^{l-m} b_{l,-m} ~.
\end{equation}
The 36 generators of Eq.~(\ref{bosgen}) can be rewritten in
angular-momentum-coupled form as
\begin{equation}
[ b^{\dagger}_{l} \times \tilde{b}_{l'} ]^{(L)}_{M} \;=\;
\sum_{mm'} \langle l,m,l',m' | L,M \rangle \,
b^{\dagger}_{lm} \tilde{b}_{l'm'} ~.
\end{equation}
The one- and two-body Hamiltonian can be expressed in terms
of the generators of $U(6)$ as
\begin{eqnarray}
H &=& \sum_l \epsilon_l \sum_m b^{\dagger}_{lm} b_{lm}
\nonumber\\
&& + \sum_L \sum_{l_1 l_2 l_3 l_4} u^{(L)}_{l_1 l_2 l_3 l_4} \,
[ [b^{\dagger}_{l_1} \times \tilde{b}_{l_2} ]^{(L)} \times
[b^{\dagger}_{l_3} \times \tilde{b}_{l_4} ]^{(L)} ]^{(0)} ~.
\label{hb}
\end{eqnarray}
In general, the Hamiltonian has to be diagonalized numerically to
obtain the energy eigenvalues and wave functions. There exist, however,
special situations in which the eigenvalues can be obtained in closed,
analytic form. These special solutions provide a framework in which
energy spectra and other nuclear properties (such as quadrupole transitions
and moments) can be interpreted in a qualitative way.
These situations correspond to dynamical symmetries of the Hamiltonian
\cite{IBM} (see section~\ref{algebraic}). 

The concept of dynamical symmetry has been shown to be a very useful tool
in different branches of physics. A well-known example in nuclear physics
is the Elliott $SU(3)$ model \cite{Elliott} to describe the properties
of light nuclei in the $sd$ shell. Another example is the $SU(3)$ flavor
symmetry of Gell-Mann and Ne'eman \cite{aGEL64} to classify the baryons
and mesons into flavor octets, decuplets and singlets and to describe
their masses with the Gell-Mann-Okubo mass formula, as described in the 
previous sections. 

The group structure
of the IBM Hamiltonian is that of $G=U(6)$. Since nuclear states have good
angular momentum, the rotation group in three dimensions $SO(3)$ should be
included in all subgroup chains of $G$ \cite{IBM}
\begin{eqnarray}
U(6)  \supset \left\{ \begin{array}{l}
U(5)  \supset SO(5) \supset SO(3) ~,\\
SU(3) \supset SO(3) ~, \\
SO(6) \supset SO(5) \supset SO(3) ~.
\end{array} \right.
\label{bchains}
\end{eqnarray}
The three dynamical symmetries which correspond to the group chains in
Eq.~(\ref{bchains}) are limiting cases of the IBM and are usually
referred to as the $U(5)$ (vibrator), the $SU(3)$ (axially symmetric rotor)
and the $SO(6)$ ($\gamma$-unstable rotor).

Here we consider a simplified form of the general expression of
the IBM Hamiltonian of Eq.~(\ref{hb}) that contains the main
features of collective motion in nuclei
\begin{equation}
H \;=\; \epsilon \, \hat n_d - \kappa \, \hat Q(\chi) \cdot \hat Q(\chi) ~,
\label{cqm}
\end{equation}
where $n_d$ counts the number of quadrupole bosons
\begin{equation}
\hat n_d \;=\; \sqrt{5} \, [d^{\dagger} \times \tilde{d} ]^{(0)}
\;=\; \sum_m d^{\dagger}_m d_m ~,
\end{equation}
and $Q$ is the quadrupole operator
\begin{equation}
\hat Q_m(\chi) \;=\; [ s^{\dagger} \times \tilde{d}
+ d^{\dagger} \times \tilde{s}
+ \chi \, d^{\dagger} \times \tilde{d} ]^{(2)}_m ~.
\end{equation}
The three dynamical symmetries are recovered for different choices
of the coefficients $\epsilon$, $\kappa$ and $\chi$. Since the IBM
Hamiltonian conserves the number of bosons and is invariant under
rotations, its eigenstates can be labeled by the total number of
bosons $N$ and the angular momentum $L$.

{\it The $U(5)$ limit.} In the absence of a quadrupole-quadrupole 
interaction $\kappa=0$,
the Hamiltonian of Eq.~(\ref{cqm}) becomes proportional to the
linear Casimir operator of $U(5)$
\begin{equation}
H_1 \;=\; \epsilon \, \hat n_d \;=\; \epsilon \, \hat{\cal C}_{1U(5)} ~.
\end{equation}
In addition to $N$ and $L$, the basis states can be labeled by
the quantum numbers $n_d$ and $\tau$, which characterize the irreducible
representations of $U(5)$ and $SO(5)$. Here $n_d$ represents the number
of quadrupole bosons and $\tau$ the boson seniority. The eigenvalues
of $H_1$ are given by the expectation value of the Casimir operator
\begin{equation}
E_1 \;=\; \epsilon \, n_d ~.
\end{equation}
In this case, the energy spectrum is characterized by a series of
multiplets, labeled by the number of quadrupole bosons, at a constant
energy spacing which is typical for a vibrational nucleus.

{\it The $SU(3)$ limit.} For the quadrupole-quadrupole interaction, 
we can distinguish two
situations in which the eigenvalue problem can be solved analytically.
If $\chi=\mp \sqrt{7}/2$, the Hamiltonian has a $SU(3)$ dynamical symmetry
\begin{equation}
H_2 \;=\; - \kappa \, \hat Q(\mp \sqrt{7}/2) \cdot
\hat Q(\mp \sqrt{7}/2) \;=\; -\frac{1}{2} \kappa
\left[ \hat{\cal C}_{2SU(3)} - \frac{3}{4} \hat{\cal C}_{2SO(3)} \right] ~.
\end{equation}
In this case, the eigenstates can be labeled by $(\lambda,\mu)$
which characterize the irreducible representations of $SU(3)$.
The eigenvalues are
\begin{equation}
E_2 \;=\; -\frac{1}{2} \kappa \left[
\lambda(\lambda+3)+\mu(\mu+3)+\lambda\mu)
-\frac{3}{4} \kappa L(L+1) \right] ~.
\end{equation}
The energy spectrum is characterized by a series of bands, in which
the energy spacing is proportional to $L(L+1)$, as in the rigid rotor
model. The ground state band has $(\lambda,\mu)=(2N,0)$ and the first
excited band $(2N-4,2)$ corresponds to a degenerate $\beta$ and $\gamma$
band. The sign of the coefficient $\chi$ is related to a prolate (-) or
an oblate (+) deformation.

{\it The $SO(6)$ limit.} For $\chi=0$, the Hamiltonian has a $SO(6)$ 
dynamical symmetry
\begin{equation}
H_3 \;=\; -\kappa \, \hat Q(0) \cdot \hat Q(0)
\;=\; -\kappa \left[ \hat{\cal C}_{2SO(6)} - \hat{\cal C}_{2SO(5)} \right] ~.
\end{equation}
The basis states are labeled by $\sigma$ and $\tau$
which characterize the irreducible representations of $SO(6)$ and $SO(5)$,
respectively. Characteristic features of the energy spectrum
\begin{equation}
E_3 \;=\; -\kappa \left[ \sigma(\sigma+4)-\tau(\tau+3) \right] ~,
\end{equation}
are the repeating patterns $L=0,2,4,2$ which is typical of the
$\gamma$-unstable rotor.

For other choices of the coefficients, the Hamiltonian of Eq.~(\ref{cqm})
describes situations in between any of the dynamical symmetries which
correspond to transitional regions, e.g. the Pt-Os isotopes exhibit a
transition between a $\gamma$-unstable and a rigid rotor
$SO(6) \leftrightarrow SU(3)$, the Sm isotopes between vibrational and
rotational nuclei $U(5) \leftrightarrow SU(3)$, and the Ru isotopes
between vibrational and $\gamma$-unstable nuclei
$U(5) \leftrightarrow SO(6)$ \cite{IBM}.

\subsection{Dynamical symmetries in odd-$A$ nuclei}

For odd-mass nuclei the IBM has been extended to include single-particle
degrees of freedom \cite{IBFM}. The Interacting Boson-Fermion Model (IBFM)
has as its building blocks a set of $N$ bosons with $l=0,2$ and an odd
nucleon (either a proton or a neutron) occupuying the single-particle
orbits with angular momenta $j=j_1,j_2,\dots$. The components of the
fermion angular momenta span the $\Omega$-dimensional space of the group
$U(\Omega)$ with $\Omega=\sum_j (2j+1)$.

One introduces, in addition to the boson creation $b^{\dagger}_i$ and
annihilation $b_i$ operators for the collective degrees of freedom,
fermion creation $a^{\dagger}_{\mu}$ and annihilation $a_{\mu}$ operators
for the single-particle. The fermion operators satisfy
anti-commutation relations
\begin{equation}
\{a_{\mu},a^{\dagger}_{\nu}\} \;=\; \delta_{\mu \nu} ~,
\hspace{1cm} \{a^{\dagger}_{\mu},a^{\dagger}_{\nu}\}
\;=\; \{a_{\mu},a_{\nu}\} \;=\; 0 ~.
\end{equation}
By construction the fermion operators commute with the boson operators.
The bilinear products
\begin{equation}
A_{\mu \nu} \;=\; a^{\dagger}_{\mu} a_{\nu} ~,
\label{fergen}
\end{equation}
generate the algebra of $U(\Omega)$, the unitary group in $\Omega$ dimensions
\begin{equation}
[ A_{\mu \nu},A_{\rho \sigma} ] \;=\; A_{\mu \sigma} \, \delta_{\nu \rho} 
- A_{\rho \nu} \, \delta_{\mu \sigma} ~.
\end{equation}
For the mixed system of boson and fermion degrees of freedom we introduce
angular-momentum-coupled generators as
\begin{equation}
B^{(L)}_M(l,l') \;=\; [ b^{\dagger}_{l} \times \tilde{b}_{l'} ]^{(L)}_{M} ~,
\hspace{1cm}
A^{(L)}_M(j,j') \;=\; [ a^{\dagger}_{j} \times \tilde{a}_{j'} ]^{(L)}_{M} ~,
\end{equation}
where $\tilde{a}_{jm}$ is defined to be a spherical tensor operator
\begin{equation}
\tilde{a}_{jm} \;=\; (-)^{j-m} a_{j,-m} ~.
\end{equation}
The most general one- and two-body rotational invariant Hamiltonian
of the IBFM can be written as
\begin{equation}
H \;=\; H_B + H_F + V_{BF} ~,
\end{equation}
where $H_B$ is the IBM Hamiltonian of Eq.~(\ref{hb}), $H_F$ is the
fermion Hamiltonian
\begin{eqnarray}
H_F &=& \sum_j \eta_j \sum_m a^{\dagger}_{jm} a_{jm}
\nonumber\\
&& + \sum_L \sum_{j_1 j_2 j_3 j_4} v^{(L)}_{j_1 j_2 j_3 j_4} \,
[ [a^{\dagger}_{j_1} \times \tilde{a}_{j_2} ]^{(L)} \times
[a^{\dagger}_{j_3} \times \tilde{a}_{j_4} ]^{(L)} ]^{(0)} ~,
\label{hf}
\end{eqnarray}
and $V_{BF}$ the boson-fermion interaction
\begin{eqnarray}
V_{BF} &=& \sum_L \sum_{l_1 l_2 j_1 j_2} w^{(L)}_{l_1 l_2 j_1 j_2} \,
[ [b^{\dagger}_{l_1} \times \tilde{b}_{l_2} ]^{(L)} \times
[a^{\dagger}_{j_1} \times \tilde{a}_{j_2} ]^{(L)} ]^{(0)} ~.
\label{vbf}
\end{eqnarray}

The IBFM Hamiltonian has an interesting algebraic structure, that suggests
the possible occurrence of dynamical symmetries in odd-$A$ nuclei.
Since in the IBFM odd-$A$ nuclei are described in terms of a mixed system
of interacting bosons and fermions, the concept of dynamical symmetries
has to be generalized. Under the restriction, that both the boson and
fermion states have good angular momentum, the respective group chains
should contain the rotation group ($SO(3)$ for bosons and $SU(2)$ for
fermions) as a subgroup
\begin{eqnarray}
U^B(6) \supset \cdots \supset SO^B(3) ~,
\nonumber\\
U^F(\Omega) \supset \cdots \supset SU^F(2) ~,
\end{eqnarray}
where we have introduced superscripts to  distinguish between boson
and fermion groups. If one of subgroups of $U^B(6)$ is isomorphic
to one of the subgroups of $U^F(\Omega)$, the boson and fermion group chains
can be combined into a common boson-fermion group chain. When the
Hamiltonian is written in terms of Casimir invariants of the combined
boson-fermion group chain, a dynamical boson-fermion symmetry arises.

{\it The $Spin(6)$ limit.} 
Among the many different possibilities, we consider two dynamical
boson-fermion symmetries associated with the $SO(6)$ limit of the IBM.
The first example discussed in the literature \cite{FI,spin6} is the case
of bosons with $SO(6)$ symmetry and the odd nucleon occupying a
single-particle orbit with spin $j=3/2$. The relevant group chains are
\begin{eqnarray}
U^B(6) \supset SO^B(6) \supset SO^B(5) \supset SO^B(3) ~,
\nonumber\\
U^F(4) \supset SU^F(4) \supset Sp^F(4) \supset SU^F(2) ~.
\end{eqnarray}
Since $SO(6)$ and $SU(4)$ are isomorphic, the boson and fermion group
chains can be combined into
\begin{eqnarray}
U^{B}(6) \otimes U^{F}(4) &\supset& SO^{B}(6) \otimes SU^{F}(4)
\nonumber\\
&\supset& Spin(6) \supset Spin(5) \supset Spin(3) ~.
\end{eqnarray}
The spinor groups $Spin(n)$ are the universal covering groups of the
orthogonal groups $SO(n)$, with $Spin(6) \sim SU(4)$, $Spin(5) \sim Sp(4)$
and $Spin(3) \sim SU(2)$. The generators of the
spinor groups consist of the sum of a boson and a fermion part.
For example, for the quadrupole operator we have
\begin{equation}
\hat Q_m \;=\; [ s^{\dagger} \times \tilde{d}
+ d^{\dagger} \times \tilde{s} ]^{(2)}_m
+ [ a^{\dagger}_{3/2} \times \tilde{a}_{3/2} ]^{(2)}_m ~.
\end{equation}
We consider a simple quadrupole-quadrupole interaction which, just as for
the $SO(6)$ limit of the IBM, can be written as the difference of two
Casimir invariants
\begin{equation}
H \;=\; -\kappa \, \hat Q \cdot \hat Q
\;=\; -\kappa \left[ \hat{\cal C}_{2Spin(6)} 
- \hat{\cal C}_{2Spin(5)} \right] ~.
\end{equation}
The basis states are classified by $(\sigma_1,\sigma_2,\sigma_3)$,
$(\tau_1,\tau_2)$ and $J$ which label the irreducible representations
of the spinor groups $Spin(6)$, $Spin(5)$ and $Spin(3)$.
The energy spectrum is obtained from the expectation value of the
Casimir invariants of the spinor groups
\begin{equation}
E \;=\; -\kappa \left[ \sigma_1(\sigma_1+4) + \sigma_2(\sigma_2+2)
+ \sigma_3^2 - \tau_1(\tau_1+3) - \tau_2(\tau_2+1) \right] ~.
\end{equation}
The mass region of the Os-Ir-Pt-Au nuclei, where the even-even Pt nuclei
are well described by the $SO(6)$ limit of the IBM and the odd proton
mainly occupies the $d_{3/2}$ shell, seems to provide
experimental examples of this symmetry, {\it e.g.} $^{191,193}$Ir and
$^{193,195}$Au \cite{FI,spin6}.

{\it The $SO(6) \otimes SU(2)$ limit.} 
The concept of dynamical boson-fermion symmetries is not restricted to
cases in which the odd nucleon occupies a single-$j$ orbit. The
first example of a multi-$j$ case discussed in the literature \cite{baha}
is that of a dynamical boson-fermion symmetry associated with the $SO(6)$
limit and the odd nucleon occupying single-particle orbits with spin
$j=1/2$, 3/2, 5/2. In this case, the fermion space is decomposed into a
pseudo-orbital part with $k=0,2$ and a pseudo-spin part with $s=1/2$
corresponding to the group reduction
\begin{eqnarray}
U^F(12) \supset U^F(6) \otimes U^F(2) \supset \left\{
\begin{array}{c} U^F(5) \otimes U^F(2) ~, \\
SU^F(3) \otimes U^F(2) ~, \\ SO^F(6) \otimes U^F(2) ~.
\end{array} \right.
\end{eqnarray}
Since the pseudo-orbital angular momentum $k$ has the same values as
the angular momentum of the $s$- and $d$- bosons of the IBM, it is clear
that the pseudo-orbital part can be combined with all three dynamical
symmetries of the IBM 
\begin{eqnarray}
U^B(6) \supset \left\{ \begin{array}{c} U^B(5) ~, \\
SU^B(3) ~, \\ SO^B(6) ~. \end{array} \right.
\end{eqnarray}
into a dynamical boson-fermion symmetry. The case, in which the bosons
have $SO(6)$ symmetry is of particular interest, since the negative
parity states in Pt with the odd neutron occupying the $3p_{1/2}$,
$3p_{3/2}$ and $3f_{5/2}$ orbits have been suggested as possible
experimental examples of a multi-$j$ boson-fermion symmetry.
In this case, the relevant boson-fermion group chain is
\begin{eqnarray}
U^{B}(6) \otimes U^{F}(12)
&\supset& U^B(6) \otimes U^{F}(6) \otimes U^{F}(2)
\nonumber\\
&\supset& U^{BF}(6) \otimes U^{F}(2)
\nonumber\\
&\supset& SO^{BF}(6) \otimes U^{F}(2)
\nonumber\\
&\supset& SO^{BF}(5) \otimes U^{F}(2)
\nonumber\\
&\supset& SO^{BF}(3) \otimes SU^{F}(2)
\nonumber\\
&\supset& SU(2) ~.
\end{eqnarray}
Just as in the first example for the spinor groups, the generators of
the boson-fermion groups consist of the sum of a boson and a fermion part,
e.g. the quadrupole operator is now written as
\begin{eqnarray}
\hat Q_m \;=\; [ s^{\dagger} \times \tilde{d}
+ d^{\dagger} \times \tilde{s} ]^{(2)}_m
&+& \sqrt{\frac{4}{5}} \, [ a^{\dagger}_{3/2} \times \tilde{a}_{1/2}
- a^{\dagger}_{1/2} \times \tilde{a}_{3/2} ]^{(2)}_m
\nonumber\\
&+& \sqrt{\frac{6}{5}} \, [ a^{\dagger}_{5/2} \times \tilde{a}_{1/2}
+ a^{\dagger}_{1/2} \times \tilde{a}_{5/2} ]^{(2)}_m ~.
\end{eqnarray}
Also in this case, the quadrupole-quadrupole interaction can be written
as the difference of two Casimir invariants
\begin{equation}
H \;=\; -\kappa \, \hat Q \cdot \hat Q
\;=\; -\kappa \left[ \hat{\cal C}_{2SO^{BF}(6)} 
- \hat{\cal C}_{2SO^{BF}(5)} \right] ~.
\end{equation}
The basis states are classified by $(\sigma_1,\sigma_2,\sigma_3)$,
$(\tau_1,\tau_2)$ and $L$ which label the irreducible representations
of the boson-fermion groups $SO^{BF}(6)$, $SO^{BF}(5)$ and $SO^{BF}(3)$.
Although the labels are the same as for the previous case, the allowed
values are different.
The total angular momentum is given by $\vec{J}=\vec{L}+\vec{s}$.
The energy spectrum is given by
\begin{equation}
E \;=\; -\kappa \left[ \sigma_1(\sigma_1+4) + \sigma_2(\sigma_2+2)
+ \sigma_3^2 - \tau_1(\tau_1+3) - \tau_2(\tau_2+1) \right] ~.
\end{equation}
The mass region of the Os-Ir-Pt-Au nuclei, where the even-even Pt nuclei
are well described by the $SO(6)$ limit of the IBM and the odd neutron
mainly occupies the negative parity orbits $3p_{1/2}$, $3p_{3/2}$ and
$3f_{5/2}$ provides experimental examples of this symmetry, in particular
the nucleus $^{195}$Pt \cite{baha,pt195,BI,sun}

\begin{figure}[t]
\centerline{\epsfig{file=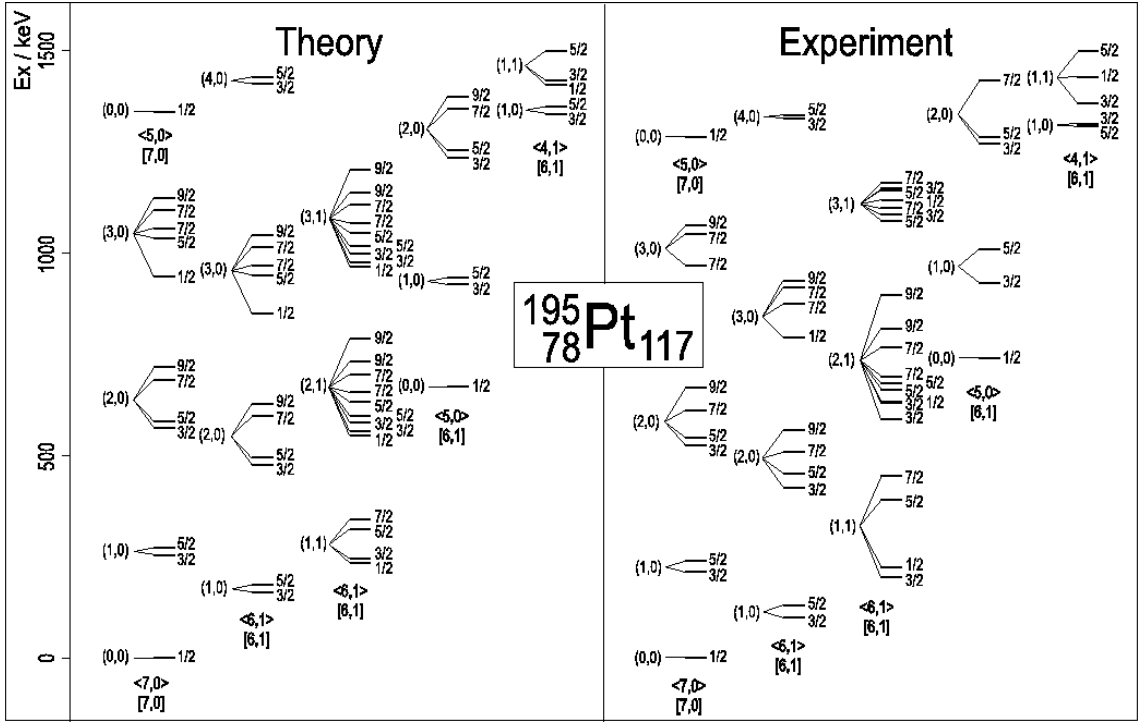,width=0.6\textwidth,angle=90}} 
\caption[]{\small Comparison between the energy spectrum of the negative
parity levels in the odd-neutron nucleus $^{195}$Pt and that obtained for the
$U(6/12)_{\nu} \otimes U(6/4)_{\pi}$ supersymmetry using Eq.~(\ref{npsusy})
with $\alpha=52.5$, $\beta=8.7$, $\gamma=-53.9$, $\delta=48.8$, 
$\epsilon=8.8$ and $\eta=4.5$ in keV.}
\label{pt195}
\end{figure}

\subsection{Dynamical supersymmetries}

Boson-fermion symmetries can further be extended by introducing the concept
of supersymmetries \cite{susy}, in which states in both even-even and
odd-even nuclei are treated in a single framework. In the previous section,
we have discussed the symmetry properties of a mixed system of boson and
fermion degrees of freedom for a fixed number of bosons $N$ and one fermion
$M=1$. The operators $B_{ij}$ and $A_{\mu \nu}$
\begin{equation}
B_{ij} = b_{i}^{\dagger} b_{j} ~, \hspace{1cm} 
A_{\mu \nu} = a_{\mu}^{\dagger} a_{\nu} ~,
\label{bgen}
\end{equation}
which generate the Lie algebra of the symmetry group 
$U^B(6) \otimes U^F(\Omega)$ of the IBFM, can only change
bosons into bosons and fermions into fermions. The number of bosons $N$ and
the number of fermions $M$ are both conserved quantities. As explained in 
Section 2.6, in addition to $B_{ij}$ and $A_{\mu \nu}$, one can introduce 
operators that change a boson into a fermion and vice versa
\begin{equation}
F_{i \mu} = b_{i}^{\dagger} a_{\mu} ~, \hspace{1cm}
G_{\mu i} = a_{\mu}^{\dagger} b_{i} ~. 
\label{fgen}
\end{equation}
The enlarged set of operators $B_{ij}$, $A_{\mu \nu}$, $F_{i \mu}$ and $G_{\mu i}$
forms a closed (super)algebra which consists of both commutation and
anticommutation relations
\begin{eqnarray}
\left[ B_{ij}, B_{kl} \right] &=&  B_{il} \delta_{jk} - B_{kj} \delta_{il} ~,
\nonumber\\
\left[ A_{\mu \nu}, A_{\rho \sigma} \right] &=& A_{\mu \sigma} \delta_{\nu \rho} 
- A_{\rho \nu} \delta_{\mu \sigma} ~,
\nonumber\\
\left[ B_{ij},A_{\mu \nu} \right] &=& 0 ~,
\nonumber\\
\left[ B_{ij},F_{k \mu} \right] &=& F_{i \mu} \delta_{jk} ~,
\nonumber\\
\left[ G_{\mu i},B_{kl} \right] &=& G_{\mu l} \delta_{ik} ~,
\nonumber\\
\left[ F_{i \mu},A_{\rho \sigma} \right] &=& F_{i \sigma} \delta_{\mu \rho} ~,
\nonumber\\
\left[ A_{\mu \nu},G_{\rho i} \right] &=& G_{\mu i} \delta_{\nu \rho} ~,
\nonumber\\
\left\{ F_{i \mu},G_{\nu j} \right\} &=& B_{ij} \delta_{\mu \nu} + A_{\nu \mu} \delta_{ij} ~,
\nonumber\\
\left\{ F_{i \mu},F_{j \nu} \right\} &=& 0 ~,
\nonumber\\
\left\{ G_{\mu i},G_{\nu j} \right\} &=& 0 ~.
\end{eqnarray}
This algebra can be identified with that of the graded Lie group $U(6/\Omega)$.
It provides an elegant scheme in which the IBM and IBFM can be unified
into a single framework \cite{susy}
\begin{equation}
U(6/\Omega) \supset U^B(6) \otimes U^F(\Omega) ~.
\end{equation}
In this supersymmetric framework,
even-even and odd-mass nuclei form the members of a supermultiplet which
is characterized by ${\cal N}=N+M$, i.e. the total number of bosons and
fermions. Supersymmetry thus distinguishes itself from ``normal'' symmetries
in that it includes, in addition to transformations among fermions and among
bosons, also transformations that change a boson into a fermion and
vice versa. 

\ba
\begin{array}{lcccccc}
\hline
& & & & & & \\
\mbox{Model} && \mbox{Generators} && \mbox{Invariant} && \mbox{Symmetry} \\
& & & & & & \\
\hline
& & & & & & \\
\mbox{IBM} &\hspace{0.3cm}& b^{\dagger}_i b_j 
&\hspace{0.3cm}& N &\hspace{0.3cm}& \mbox{U(6)} \\
& & & & & & \\
\mbox{IBFM} && b^{\dagger}_i b_j ~,\; a^{\dagger}_{\mu} a_{\nu} && N, M && 
\mbox{U(6)} \otimes \mbox{U($\Omega$)} \\
& & & & & & \\
\mbox{n-SUSY} && b^{\dagger}_i b_j ~,\; a^{\dagger}_{\mu} a_{\nu} ~,\; 
b^{\dagger}_i a_{\mu}  ~,\; a^{\dagger}_{\mu} b_i && {\cal N} 
&& \mbox{U(6/$\Omega$)} \\
& & & & & & \\
\hline
\end{array}
\nonumber
\ea

The Os-Ir-Pt-Au mass region provides ample
experimental evidence for the occurrence of dynamical (super)symmetries in
nuclei. The even-even nuclei $^{194,196}$Pt are the standard examples of
the $SO(6)$ limit of the IBM \cite{so6} and the odd proton, in first
approximation, occupies the single-particle level $2d_{3/2}$. In this
special case, the boson and fermion groups can be combined into spinor groups,
and the odd-proton nuclei $^{191,193}$Ir and $^{193,195}$Au were suggested
as examples of the $Spin(6)$ limit \cite{FI,spin6}. The appropriate extension
to a supersymmetry is by means of the graded Lie group $U(6/4)$
\begin{eqnarray}
U(6/4) \supset U^{B}(6) \otimes U^{F}(4)
&\supset& SO^{B}(6) \otimes SU^{F}(4)
\nonumber\\
&\supset& Spin(6) \supset Spin(5) \supset Spin(3) \supset Spin(2) ~.
\end{eqnarray}
The pairs of nuclei $^{190}$Os - $^{191}$Ir, $^{192}$Os - $^{193}$Ir,
$^{192}$Pt - $^{193}$Au and $^{194}$Pt - $^{195}$Au have been analyzed as
examples of a $U(6/4)$ supersymmetry \cite{susy}.

Another example of a dynamical supersymmetry in this mass region
is that of the Pt nuclei. The even-even isotopes are well described
by the $SO(6)$ limit of the IBM and the odd neutron mainly occupies
the negative parity orbits $3p_{1/2}$, $3p_{3/2}$ and $3f_{5/2}$.
In this case, the graded Lie group is $U(6/12)$
\begin{eqnarray}
U(6/12) \supset U^{B}(6) \otimes U^{F}(12)
&\supset& U^B(6) \otimes U^{F}(6) \otimes U^{F}(2)
\nonumber\\
&\supset& U^{BF}(6) \otimes U^{F}(2)
\nonumber\\
&\supset& SO^{BF}(6) \otimes U^{F}(2)
\nonumber\\
&\supset& SO^{BF}(5) \otimes U^{F}(2)
\nonumber\\
&\supset& SO^{BF}(3) \otimes SU^{F}(2)
\nonumber\\
&\supset& SU(2) ~.
\end{eqnarray}
The odd-neutron nucleus $^{195}$Pt, together with $^{194}$Pt, were
studied as an example of a $U(6/12)$ supersymmetry \cite{baha,BI,sun}.

\subsection{Dynamical neutron-proton supersymmetries}

As we have seen in the previous section, the mass region $A \sim 190$ has
been a rich source of possible empirical evidence for the existence of
(super)symmetries in nuclei. The pairs of nuclei $^{190}$Os - $^{191}$Ir,
$^{192}$Os - $^{193}$Ir, $^{192}$Pt - $^{193}$Au and $^{194}$Pt - $^{195}$Au
have been analyzed as examples of a $U(6/4)$ supersymmetry \cite{susy},
and the nuclei $^{194}$Pt - $^{195}$Pt as an example of a $U(6/12)$
supersymmetry \cite{baha}.
These ideas were later extended to the case where neutron and proton bosons
are distinguished \cite{quartet}, predicting in this way a correlation among
quartets of nuclei, consisting of an even-even, an odd-proton, an odd-neutron
and an odd-odd nucleus. The best experimental example of such a
quartet with $U(6/12)_{\nu} \otimes U(6/4)_{\pi}$ supersymmetry is provided
by the nuclei $^{194}$Pt, $^{195}$Au, $^{195}$Pt and $^{196}$Au. 
\begin{eqnarray}
\begin{array}{ccccc}
& ^{194}_{ 78}\mbox{Pt}_{116} & \leftrightarrow
& ^{195}_{ 78}\mbox{Pt}_{117} & \\
& & & & \\
& \updownarrow & & \updownarrow & \\
& & & & \\
& ^{195}_{ 79}\mbox{Au}_{116} & \hspace{1cm} \leftrightarrow \hspace{1cm}
& ^{196}_{ 79}\mbox{Au}_{117} & \\
\end{array}
\label{magic}
\end{eqnarray}
The number of bosons and fermions are related to the number of valence nucleons, 
{\it i.e.} the number of protons and neutrons outside the closed shells. The 
relevant closed shells are $Z=82$ for protons and $N=126$ for neutrons. For the 
even-even nucleus $^{194}_{ 78}\mbox{Pt}_{116}$ the number of bosons are 
$N_{\pi}=(82-78)/2=2$ and $N_{\nu}=(126-116)/2=5$. There are no unpaired nucleons 
$M_{\pi}=M_{\nu}=0$. For the odd-neutron nucleus $^{195}_{ 78}\mbox{Pt}_{117}$  
there are 9 valence neutrons which leads to $N_{\nu}=4$ neutron bosons and 
$M_{\nu}=1$ unpaired neutron. The $_{79}\mbox{Au}$ isotopes have 3 valence 
protons which are divided over $N_{\pi}=1$ proton boson and 
$M_{\pi}=1$ unpaired proton. This supersymmetric quartet of nuclei is characterized 
by ${\cal N}_{\pi}=N_{\pi}+M_{\pi}=2$ and ${\cal N}_{\nu}=N_{\nu}+M_{\nu}=5$.
The number of bosons and fermions are summarized in Table~\ref{number}. 

\begin{table}[h]
\centering
\caption[]{\small The number of bosons and fermions of a 
supersymmetric quartet of nuclei.}
\label{number}
\vspace{15pt}
\begin{tabular}{ccccc}
\hline 
& & & & \\
Nucleus & $N_{\pi}$ & $M_{\pi}$ & $N_{\nu}$ & $M_{\nu}$ \\
& & & & \\
\hline
& & & & \\
$^{194}_{ 78}\mbox{Pt}_{116}$ & 2 & 0 & 5 & 0 \\
& & & & \\
$^{195}_{ 78}\mbox{Pt}_{117}$ & 2 & 0 & 4 & 1 \\
& & & & \\
$^{195}_{ 79}\mbox{Au}_{116}$ & 1 & 1 & 5 & 0 \\
& & & & \\
$^{196}_{ 79}\mbox{Au}_{117}$ & 1 & 1 & 4 & 1 \\
& & & & \\
\hline
\end{tabular}
\end{table}

\begin{figure}[t]
\centerline{\epsfig{file=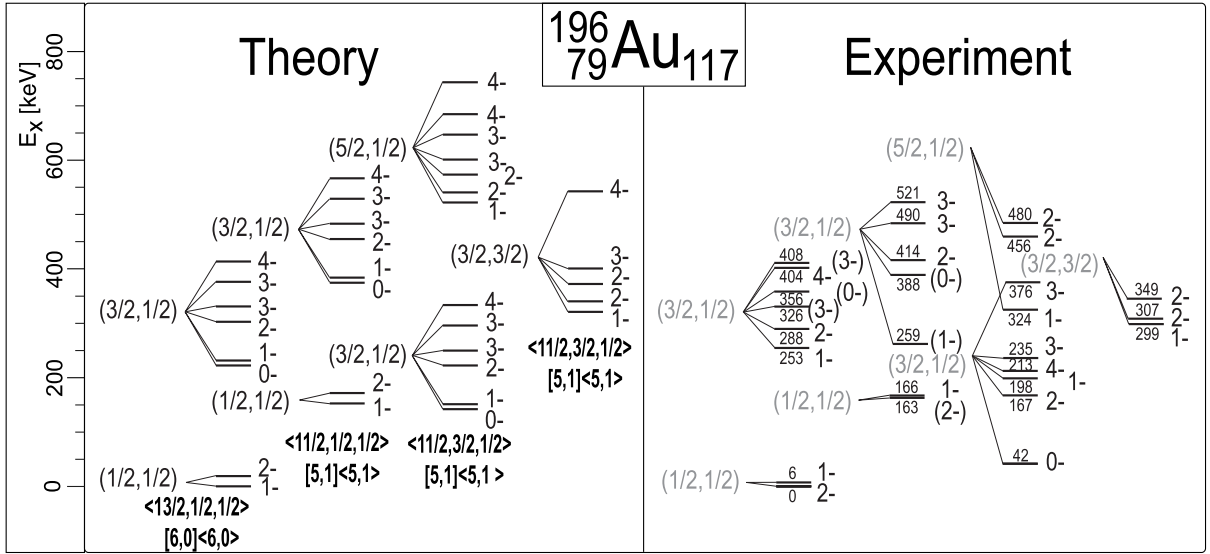,width=\textwidth}} 
\caption[]{\small As Fig.~\ref{pt195}, but for the negative
parity levels in the odd-odd nucleus $^{196}$Au.}
\label{au196}
\end{figure}

In previous sections, we have used a schematic Hamiltonian consisting
only of a quadrupole-quadrupole interaction to discuss the different
dynamical symmetries. In general, a dynamical (super)symmetry arises
whenever the Hamiltonian is expressed in terms of the Casimir invariants
of the subgroups in a group chain. The relevant subgroup chain of
$U(6/12)_{\nu} \otimes U(6/4)_{\pi}$ for the Pt and Au
nuclei is given by \cite{quartet}
\begin{eqnarray}
U(6/12)_{\nu} \otimes U(6/4)_{\pi} &\supset&
U^{B_{\nu}}(6) \otimes U^{F_{\nu}}(12) \otimes
U^{B_{\pi}}(6) \otimes U^{F_{\pi}}(4)
\nonumber\\
&\supset& U^B(6) \otimes U^{F_{\nu}}(6) \otimes U^{F_{\nu}}(2) \otimes
U^{F_{\pi}}(4)
\nonumber\\
&\supset& U^{BF_{\nu}}(6) \otimes U^{F_{\nu}}(2) \otimes U^{F_{\pi}}(4)
\nonumber\\
&\supset& SO^{BF_{\nu}}(6) \otimes U^{F_{\nu}}(2) \otimes SU^{F_{\pi}}(4)
\nonumber\\
&\supset& Spin(6) \otimes U^{F_{\nu}}(2)
\nonumber\\
&\supset& Spin(5) \otimes U^{F_{\nu}}(2)
\nonumber\\
&\supset& Spin(3) \otimes SU^{F_{\nu}}(2)
\nonumber\\
&\supset& SU(2) ~.
\end{eqnarray}
In this case, the Hamiltonian
\begin{eqnarray}
H &=& \alpha \, \hat{\cal C}_{2U^{BF_{\nu}}(6)} 
+ \beta \, \hat{\cal C}_{2SO^{BF_{\nu}}(6)}
+ \gamma \, \hat{\cal C}_{2Spin(6)}
\nonumber\\
&& + \delta \, \hat{\cal C}_{2Spin(5)} 
+ \epsilon \, \hat{\cal C}_{2Spin(3)}
+ \eta \, \hat{\cal C}_{2SU(2)} ~,
\end{eqnarray}
describes simultaneously the excitation spectra of the quartet of nuclei.
Here we have neglected terms that only contribute to binding energies.
The energy spectrum is given by the eigenvalues of the Casimir operators
\begin{eqnarray}
E &=& \alpha \, \left[ N_1(N_1+5) + N_2(N_2+3) + N_3(N_3+1) \right]
\nonumber\\
&& + \beta \, \left[ \Sigma_1(\Sigma_1+4) + \Sigma_2(\Sigma_2+2)
+ \Sigma_3^2 \right]
\nonumber\\
&& + \gamma \, \left[ \sigma_1(\sigma_1+4) + \sigma_2(\sigma_2+2)
+ \sigma_3^2 \right]
\nonumber\\
&& + \delta \, \left[ \tau_1(\tau_1+3) + \tau_2(\tau_2+1) \right]
+ \epsilon \, J(J+1) + \eta \, L(L+1) ~.
\label{npsusy}
\end{eqnarray}
The coefficients $\alpha$, $\beta$, $\gamma$, $\delta$, $\epsilon$ and
$\eta$ have been determined in a simultaneous fit of the excitation energies
of the four nuclei of Eq.~(\ref{magic}) \cite{au196}.

The supersymmetric classification of nuclear levels in the Pt
and Au isotopes has been re-examined by taking advantage of the significant
improvements in experimental capabilities developed in the last decade.
High resolution transfer experiments with protons and polarized deuterons
have strengthened the evidence for the existence of supersymmetry in
atomic nuclei. The experiments include high resolution transfer experiments
to $^{196}$Au at TU/LMU M\"unchen \cite{tres,pt195}, and in-beam gamma ray
and conversion electron spectroscopy following the reactions
$^{196}$Pt$(d,2n)$ and $^{196}$Pt$(p,n)$ at the cyclotrons of the PSI and
Bonn \cite{au196}. These studies have achieved an improved classification
of states in $^{195}$Pt and $^{196}$Au which give further support to the
original ideas \cite{FI,baha,quartet} and extend and refine previous
experimental work in this research area.

In dynamical (super)symmetries closed expressions can be derived for energies, 
as well as selection rules and intensities for electromagnetic transitions and
transfer reactions. Recent work in this area concerns a study of one- and 
two-nucleon transfer reactions: 
\begin{itemize}
\item As a consequence of the supersymmetry, explicit correlations 
were found between the spectroscopic factors of the one-proton 
reactions between n-SUSY partners  
$^{194}\mbox{Pt} \leftrightarrow ^{195}\mbox{Au}$ and 
$^{195}\mbox{Pt} \leftrightarrow ^{196}\mbox{Au}$ \cite{BBF1}
which can be tested experimentally. 
\item The spectroscopic stengths of two-nucleon transfer reactions constitute a 
stringent test for two-nucleon correlations in the nuclear wave functions. 
A study in the framework of nuclear supersymmetry led to a set of closed 
analytic expressions for ratios of spectroscopic factors. Since these ratios 
are parameter independent they provide a direct test of the wave functions. 
A comparison between the recently measured 
$^{198}\mbox{Hg}(\vec{d},\alpha)^{196}\mbox{Au}$ reaction \cite{Wirth} and the 
predictions of the nuclear quartet supersymmetry \cite{BBF2} lends further 
support to the validity of supersymmetry in nuclear physics.   
\end{itemize}

\section{Prehispanic supersymmetry}

\begin{figure}[b]
\centerline{\epsfig{file=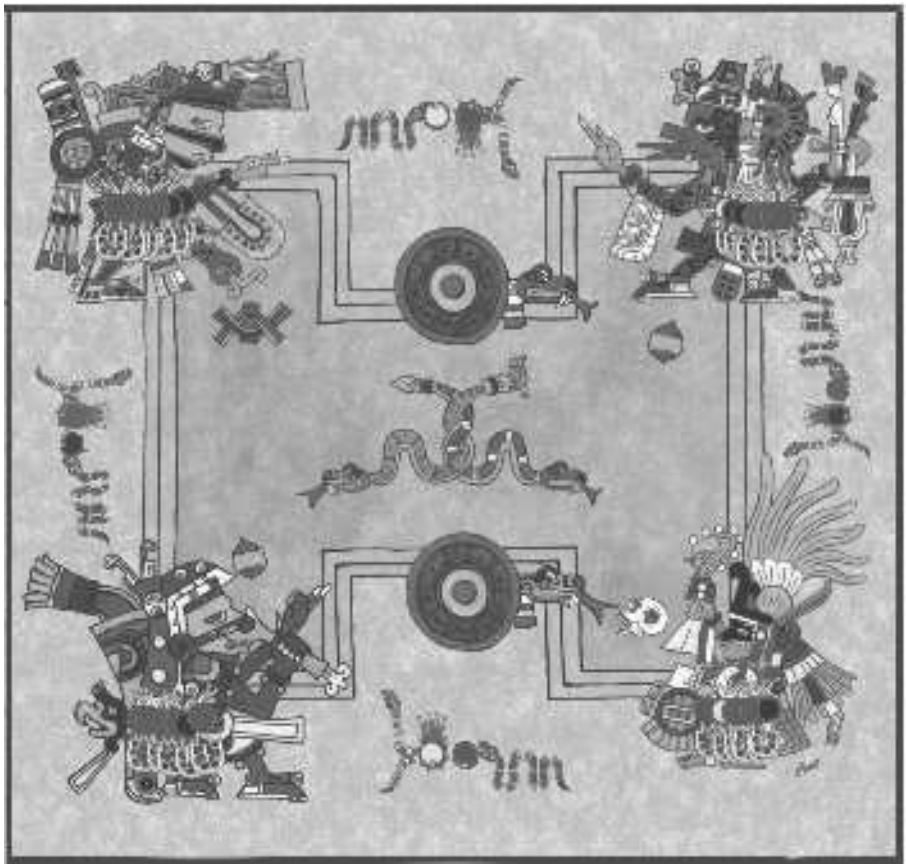,width=0.75\textwidth}} 
\caption[]{\small Prehispanic supersymmetry}
\label{aztec}
\end{figure}

Fig.~\ref{aztec} shows an artistic interpretation of supersymmetry in physics. 
This figure is part of the design of the poster of the {\it XXXV Latin-American 
School of Physics. Supersymmetries in Physics and its Applications} (ELAF 2004) 
by Renato Lemus which is inspired by the concept 
of supersymmetry as used in nuclear and particle physics and the 
`Juego de Pelota', the ritual game of prehispanic cultures of Mexico.
The four players on the ballcourt are aztec gods which represent 
the nuclei of a supersymmetric quartet. Each one of the gods   
represents a nucleus, on the 
top left {\it Tezcatlipoca}: the even-even nucleus $^{194}$Pt, 
top right {\it Quetzalc\'oatl}: the odd-even nucleus $^{195}$Pt, 
bottom left {\it Camaxtle}: the even-odd nucleus $^{195}$Au, 
and finally {\it Huitzilopochtli}: the odd-odd nucleus $^{196}$Au. 
The association between the gods and the nuclei is made via the 
number and the color of the balls that each one of the players carry. 
Each player carries 7 balls. 
The green and blue balls correspond to the neutron and proton bosons, 
whereas the yellow and red ones correspond to neutrons and protons, 
respectively. The one-nucleon transfer operators that induce 
the supersymmetric transformation between different nuclei, are 
represented by red coral snakes (`coralillos'). The snakes that create 
a particle carry a ball in their mouth whose color indicates the type 
of particle. On the other hand, the snakes that annihilate a particle 
carry the corresponding ball soaking with blood that seems to 
split their body. Both types of snakes we see in segmented form, in 
representation of the quantization of energy. In the world of the 
ancient Mexico both living and dead creatures form a coherent unity 
and harmonize in the same plane of importance. This is reflected in 
the eyes that are included in all components of a graphical representation. 
For this reason, the balls associated with the creation and annihilation 
of particles have eyes. 

The central figure in the ball court consists of two intertwined snakes, 
a coral snake and a rattle snake. They represent another aspect of 
supersymmetry as it is used in particle physics, in which each particle has 
its supersymmetric counterpart. The reason that this is symbolized 
by snakes is their property to change skins. Thus, a change of skin of 
two apparently different snakes suggests the transformation between bosons 
and fermions. The same two snakes make their appearance on the circular 
stone rings, the `score board' of the aztec ball game. 
In the ball court one finds, at the feet of {\it Tezcatlipoca} the 
symbol {\it Ollin}, movement, which represents the uncertainty principle. 
Similarly, we see a heart in the upper right and the lower left part. 
The hearts have two meanings. On the one hand they characterize the ritual 
aspect of the ancient game `Juego de Pelota' and, on the other hand,
they represent the `road with a heart', which science could follow. 
Finally, next to {\it Huitzilopochtli} there is a skull to remind us of 
the fleeting nature of our existence. 

More information can be found in the proceedings of the ELAF 2004 \cite{elaf}. 

\section{Summary and conclusions}

The concept of symmetry has played a very important role in physics, 
especially in the 20th century with the development of quantum mechanics 
and quantum field theory. The applications involve among other 
geometric symmetries, permutation symmetries, space-time symmetries, 
gauge symmetries and dynamical symmetries. In these lecture notes, 
I have concentrated mainly on the latter. The basic idea of dynamical 
symmetries is that of finding order, regularity and simple patterns  
in complex many-body systems. The examples discussed in these notes 
include isospin and flavor symmetry and nuclear supersymmetry.   

Dynamical symmetries not only provide classification schemes for 
finite quantal systems and simple benchmarks against which the 
experimental data can be interpreted in a clear and transparent 
manner, but also led to important predictions that have been 
verified later experimentally, such as the $\Omega^-$ baryon as the 
missing member of the baryon decuplet, the nucleus $^{196}$Pt as an 
example of the $SO(6)$ limit of the IBM and the odd-odd nucleus $^{196}$Au 
whose spectroscopic properties had been predicted as a 
consequence of nuclear supersymmetry almost 15 years before they 
were measured. 

The interplay between theory and experiment is reflected in 
the combination of the Platonic ideal of symmetry with
the more down-to-earth Aristotelic ability to recognize complex
patterns in Nature.

\begin{figure}[t]
\centerline{\epsfig{file=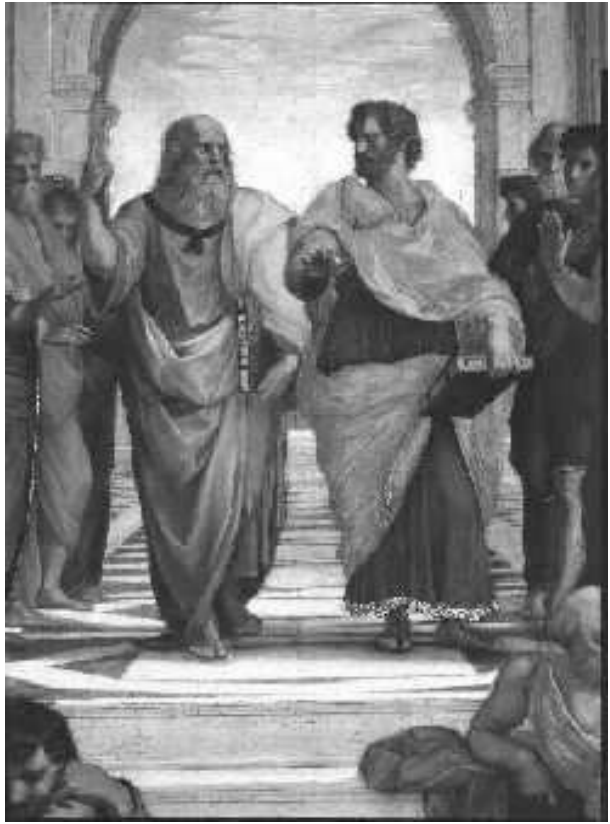,width=0.5\textwidth}} 
\caption{\small Detail of ``The School of Athens'' (Plato on the left and
Aristoteles on the right), by Rafael.}
\label{athens}
\end{figure}

\section*{Acknowledgments}

It is a pleasure to thank Jos\'e Barea and Alejandro Frank 
for many interesting discussions on supersymmetric nuclei. 
This paper was supported in part by Conacyt, Mexico.


\begin{thebibliography}{99}

\bibitem{MacTutor}
See {\it e.g.} MacTutor History of Mathematics archive, 
School of Mathematics and Statistics, University of St. Andrews, Scotland 
http://www-history.mcs.st-andrews.ac.uk/history

\bibitem{Iachello}
F. Iachello, Nucl. Phys. A {\bf 751}, 329c (2005). 

\bibitem{Pauli}
W. Pauli, Z. Physik {\bf 36}, 336 (1926).

\bibitem{Piet}
P. Van Isacker, Rep. Prog. Phys. {\bf 62}, 1661 (1999);
Nucl. Phys. A {\bf 704}, 232c (2002). 

\bibitem{oromana}
A. Frank, J. Barea and R. Bijker, 
in {\it The Hispalensis Lectures on Nuclear Physics, Vol. 2}, 
Eds. J.M. Arias and M. Lozano, 
Lecture Notes in Physics {\bf 652} (2004), 285-324 [arXiv:nucl-th/0402058].

\bibitem{hamermesh}
M. Hamermesh, 
{\it Group Theory and its Applications to Physical Problems}, 
(Addison Wesley, Reading, 1962  
and Dover Publications, New York, 1989).

\bibitem{lipkin}
H.J. Lipkin,
{\it Lie Groups for Pedestrians},
(North-Holland, Amsterdam, 1966 
and Dover Publications, New York, 2002).

\bibitem{gilmore}
R. Gilmore,
{\it Lie Groups, Lie Algebras, and Some of Their Applications},
(Wiley-Interscience, New York, 1974).

\bibitem{wybourne}
B.G. Wybourne,
{\it Classical Groups for Physicists},
(Wiley-Interscience, New York, 1974).

\bibitem{elliott}
J.P. Elliott and P.G. Dawber, 
{\it Symmetry in Physics}, 
(Oxford University Press, Oxford, 1979).

\bibitem{georgi}
H. Georgi, 
{\it Lie Algebras in Particle Physics: from Isospin to Unified Theories}, 
(Addison Wesley, 1982)

\bibitem{stancu}
Fl. Stancu, 
{\it Group Theory in Subnuclear Physics}, 
(Oxford University Press, Oxford, 1996)
 
\bibitem{hill}
E.L. Hill, Rev. Mod. Phys. {\bf 23}, 253 (1951).

\bibitem{aHEI32}
W. Heisenberg,
Z. Phys. {\bf 77}, 1 (1932).

\bibitem{aBOH75}
A. Bohr and B.R. Mottelson,
{\it Nuclear Structure. II. Nuclear Deformations}
(Benjamin, New York, 1975).

\bibitem{aGEL62}
M. Gell-Mann,
Phys. Rev. {\bf 125}, 1067 (1962).

\bibitem{aOKU62}
S. Okubo,
Progr. Theor. Phys. {\bf 27}, 949 (1962).

\bibitem{aGEL64}
M. Gell-Mann and Y. Ne'eman,
{\it The Eightfold Way},
(Benjamin, New York, 1964).

\bibitem{IBM}
F. Iachello and A. Arima,
{\it The Interacting Boson Model},
(Cambridge University Press, Cambridge, 1987).

\bibitem{IBFM}
F. Iachello and P. Van Isacker,
{\it The Interacting Boson-Fermion Model}
(Cambridge University Press, Cambridge, 1991).

\bibitem{FI}
F. Iachello,
Phys. Rev. Lett. {\bf 44}, 772 (1980).

\bibitem{susy}
A.B. Balantekin, I. Bars and F. Iachello,
Phys. Rev. Lett. {\bf 47}, 19 (1981);\\
A.B. Balantekin, I. Bars and F. Iachello,
Nucl. Phys. A {\bf 370}, 284 (1981).

\bibitem{baha}
A.B. Balantekin, I. Bars, R. Bijker and F. Iachello,
Phys. Rev. C {\bf 27}, 1761 (1983). 

\bibitem{thesis}
R. Bijker, Ph.D. Thesis, University of Groningen (1984).

\bibitem{tres}
A. Metz, J. Jolie, G. Graw, R. Hertenberger, J. Gr\"oger,
C. G\"unther, N. Warr and Y. Eisermann,
Phys. Rev. Lett. {\bf 83}, 1542 (1999).

\bibitem{pt195}
A. Metz, Y. Eisermann, A. Gollwitzer, R. Hertenberger, B.D. Valnion,
G. Graw and J. Jolie,
Phys. Rev. C {\bf 61}, 064313 (2000).

\bibitem{au196}
J. Gr\"oger, J. Jolie, R. Kr\"ucken, C.W. Beausang, M. Caprio, R.F. Casten,
J. Cederkall, J.R. Cooper, F. Corminboeuf, L. Genilloud, G. Graw,
C. G\"unther, M. de Huu, A.I. Levon, A. Metz, J.R. Novak, N. Warr and
T. Wendel,
Phys. Rev. C {\bf 62}, 064304 (2000).

\bibitem{quartet}
P. Van Isacker, J. Jolie, K. Heyde and A. Frank,
Phys. Rev. Lett. {\bf 54}, 653 (1985).

\bibitem{Elliott}
J.P. Elliott,
Proc. Roy. Soc. A {\bf 245}, 128 (1958);
{\it ibid.} {\bf 245}, 562 (1958).

\bibitem{spin6}
F. Iachello and S. Kuyucak,
Ann. Phys. (N.Y.) {\bf 136}, 19 (1981).

\bibitem{BI}
R. Bijker and F. Iachello,
Ann. Phys. (N.Y.) {\bf 161}, 360 (1985).

\bibitem{sun}
H.Z. Sun, A. Frank and P. Van Isacker,
Phys. Rev. C {\bf 27}, 2430 (1983);\\
H.Z. Sun, A. Frank and P. Van Isacker,
Ann. Phys. (N.Y.) {\bf 157}, 183 (1984).

\bibitem{so6}
J.A. Cizewski, R.F. Casten, G.J. Smith, M.L. Stelts, W.R. Kane,
H.G. B\"orner and W.F. Davidson,
Phys. Rev. Lett. {\bf 40}, 167 (1978);\\
A. Arima and F. Iachello,
Phys. Rev. Lett. {\bf 40}, 385 (1978).

\bibitem{BBF1}
J. Barea, R. Bijker and A. Frank, 
J. Phys. A: Math. Gen. {\bf 37}, 10251 (2004).

\bibitem{Wirth}
H.-F. Wirth, G. Graw, S. Christen, Y. Eisermann, A. Gollwitzer, 
R. Hertenberger, J. Jolie, A. Metz, O. M\"oller, D. Tonev and B.D. Valnion, 
Phys. Rev. C {\bf 70}, 014610 (2004).

\bibitem{BBF2}
J. Barea, R. Bijker and A. Frank, 
Phys. Rev. Lett. {\bf 94}, 152501 (2005).

\bibitem{elaf}
R. Lemus, 
in {\it Latin-American School of Physics XXXV ELAF. 
Supersymmetries in Physics and its Applications},  
Eds. R. Bijker et al., 
AIP Conference Proceedings {\bf 744} (2005), xi-xvi. 

\end{thebibliography}
\end{document}